# Cosmological pressure fluctuations and spatial expansion

Dale R. Koehler
(Sandia Park, New Mexico)

Abstract

Most recently, experimental determinations of the spectrometric characteristics and internal structural velocities of galaxies have suggested the presence of massive central black holes. The analyses of the galactic spectrometric electromagnetic frequency shifts have resulted in a correlation between the hole mass and the surrounding bulge mass. In the present work, we examine whether conditions existed in the early universe, that could have led to the formation of gravitational structures possessing such unusual characteristics. We propose an early-time pressure-fluctuation model, which would have generated a radiation based energy distribution possessing the characteristic of a centrally collapsed zone isolated from its surrounding environment and thereby manifesting such a black hole behavior**.** As a hole-core expansion model, it exhibits a time evolving matter and radiation distribution, leading to a supplementary treatment of early time cosmological energy fluctuations. Einstein's gravitational equations are assumed to apply within the radiation-dominated hole-core spatial domain and, with utilization of a spherically symmetric isotropic metric, are used in order to calculate the evolutionary time expansion characteristics. Birth times for the radiation structures are uniquely correlated with the size of the spheres and are primarily determined from the early time energy densities and the apparent curvatures presented by the gravitational equations. Pressure and temperature characteristics are calculated. The hole-core model is described as a flat metric, matter plus radiation, $\sigma = 1/3$, energy distribution. It displays an early time pressure fluctuation collapse, tentatively interpreted to be the formation of a galaxy hole, and therein provides a theoretical basis for the experimental data.

1. INTRODUCTION

In the early part of the century, Hubble and Humason [1] cataloged the recessional velocities, relative to our Milky Way galaxy, of numerous galaxies, thereby identifying the velocity versus distance relationship now known as the Hubble law. Recessional velocities were calculable from analyses of the spectroscopic data associated with the individual light sources while distances were determined from the apparent magnitudes of these sources. It is now believed that these optical Doppler frequency shifts arise from a cosmological expansion of the intervening space itself, Einstein [2], rather than from a motion of the galaxies through space, Silk [3], Ferris [4]. More recently the experimental measurements have been extended to examine the internal galactic structure with results that suggest the presence of a massive black hole central to most, if not all, observable galaxies, Kormendy [5], Magorrian [6], Kormendy [7] and Gebhardt [8].

Did conditions exist in the early universe which could have led to the formation of gravitational structures possessing such unusual characteristics? In the present work, we examine an early-time pressure-fluctuation model, which would have generated a radiation based energy distribution possessing the characteristic of a centrally collapsed zone isolated from its surrounding environment and thereby exhibiting such a black hole behavior. To further explore this description, pressure fluctuations with energy densities simulating experimentally observed hole and bulge masses have been investigated. The galaxy-core model, as developed, leads to energy densities and formation times when radiation was the dominant cosmological energy contributor. The birth, or formation, times are less than a year, even for the largest energy considered which was equivalent to approximately $10^{12}$ solar masses. At these early times the radiation energy dominates by a factor of several orders of magnitude over matter energy densities. Since we are considering structural birth times in the range of $(10)^{-4}$ to $(10)^{-1}$ years, cosmological expansion factors are in the range of $(10)^{-8}$ to $(10)^{-6}$ with temperatures in the range of $(10)^{8}$ K and radiation densities of the order of $(10)^{2}$ to $(10)^{-3}$ grams/cm$^{3}$.

## 2. CORE AND HOLE MODELING

The galaxy-core model begins with an Einsteinian gravitational treatment of the early matter and radiation density fluctuation defining the core. Einstein's gravitational equations, Einstein [9], when qualified on a cosmological scale to an isotropic homogeneous space, have been solved by Friedmann, Einstein [9], to yield the classic expanding universe equations for a perfect fluid, or continuous matter distribution;

$$G_{ab}(g_{ab}) \equiv R_{ab} - \frac{1}{2}g_{ab}R = 8\pi T_{ab} = 8\pi[\rho u_a u_b + p(g_{ab} + u_a u_b)] \quad . \quad (1)$$

$G_{ab}$ is the Einstein tensor, a function of the metric $g_{ab}$ and its first two derivatives, $R_{ab}$ is the Ricci tensor, $R$ the Ricci scalar, and $T_{ab}$ the energy-momentum tensor describing the material contents of the environment; $\rho$ = total energy density and $p = constant\ x\ \rho = \sigma\rho$ = pressure energy density. The commonly used Robertson-Walker metric, for the three curved spaces is

$$ds^2 = -a(t)^2\left(1+\frac{kr^2}{4}\right)^{-2}\left[dr^2 + r^2 d\Omega^2\right] + dt^2 \,,\ where$$
$$k = +1, 0, -1, \tag{2}$$

for a sphere, plane or pseudosphere. The Friedmann solutions to these equations are

$$\frac{\ddot{a}}{a} = -\frac{4\pi}{3}(\rho+3p) \quad and \quad \left(\frac{\dot{a}}{a}\right)^2 = \frac{8\pi}{3}\rho - \frac{k}{a^2}\,, \tag{3}$$

where $a$ is the time evolving spatial expansion parameter and $k$ is the space curvature parameter.

As a similar model for the present calculations, the galaxy-core has been described in evolutionary terms as a continuous matter and radiation density fluctuation with $\sigma = 1/3$, with a flat space metric (producing an associated radius dependent attraction or curvature), and beginning at time $t_{core-birth}$. This is an evolving spatial and matter region, which subsequently begins collapsing in the central core zone, a temporal diminution of the hole. However, describing this spherically symmetric, radial coordinate centered, pressure fluctuation region requires a modified metric, which in the present treatment employs the isotropic form utilized by Tolman [10],

$$ds^2 = g_{11}\left[dr^2 + r^2 d\Omega^2\right] + g_{44} dt^2 = $$
$$= -e^{\mu}\left[dr^2 + r^2 d\Omega^2\right] + e^{\nu} dt^2 \quad, \tag{4}$$

where
$$\mu = \mu(r,t) \quad and \quad \nu = \nu(r,t) \quad.$$

The system of equations represented by equation (1) and as calculated by Tolman [10], is shown more explicitly in equation (5);

$$8\pi T_1^1 = -e^{-\mu}\left[\frac{\mu'^2}{4} + \frac{\mu'\nu'}{2} + \frac{\mu'+\nu'}{r}\right] + e^{-\nu}\left[\ddot{\mu} + \frac{3}{4}\dot{\mu}^2 - \frac{\dot{\mu}\dot{\nu}}{2}\right],$$

$$8\pi T_2^2 = -e^{-\mu}\left[\frac{\mu''}{2} + \frac{\nu''}{2} + \frac{\nu'^2}{4} + \frac{\mu'+\nu'}{2r}\right] + e^{-\nu}\left[\ddot{\mu} + \frac{3}{4}\dot{\mu}^2 - \frac{\dot{\mu}\dot{\nu}}{2}\right] = 8\pi T_3^3 \;,$$

$$8\pi T_4^4 = -e^{-\mu}\left[\mu'' + \frac{\mu'^2}{4} + \frac{2\mu'}{r}\right] + e^{-\nu}\left[\frac{3}{4}\dot{\mu}^2\right],$$

$$8\pi T_4^1 = +e^{-\mu}\left[\dot{\mu}' - \frac{\dot{\mu}\nu'}{2}\right],$$

$$8\pi T_1^4 = -e^{-\mu}\left[\dot{\mu}' - \frac{\dot{\mu}\nu'}{2}\right]. \tag{5}$$

We restrict the present modeling to the case of no radial energy flow thereby requiring that

$$\dot{\mu}' = \frac{\dot{\mu}\nu'}{2}. \tag{6}$$

The static solutions and a general dynamic solution to this equation are

1. $\dot{\mu} = 0$,
2. $\mu = f(r) + g(t)$ ; $\nu = \nu(t)$ or $const$, and
3. $\mu(r,t) = \mu_0 e^{\nu(r,t)/2} + \mu_1$ ; $\nu(r,t) = \varepsilon(r) + \lambda(t)$. (7)

The usual notation, where primes denote differentiation with respect to the radial coordinate $r$ and dots denote differentiation with respect to the time coordinate $t$, is employed. Static solution 1, Schwarzschild's interior and exterior solutions for the case of an incompressible perfect fluid sphere of constant density surrounded by empty space, is provided in Tolman [10]. A zero-pressure surface-condition and matching and normalization of the interior and exterior metrics at the sphere radius were used as boundary conditions. Here we use the general solution form 3 of equation (7),

$$\mu(r,t) = \mu_0 e^{\nu(r,t)/2} + \mu_1,$$
$$\nu(r,t) = \varepsilon(r) + \lambda(t). \quad (8)$$

$\mu_1$ is either a constant or a function of $r$, that is $\mu_1(r)$. The radiation, or perfect fluid, character of the model requires that

$$T_1^1 = T_2^2 \quad and$$
$$T_1^1 = -\frac{1}{3} T_4^4. \quad (9)$$

From equation set (5) then,

$$-e^{-\mu}\left[\frac{\mu'^2}{4} + \frac{\mu'\nu'}{2} + \frac{\mu'+\nu'}{r}\right] = -e^{-\mu}\left[\frac{\mu''}{2} + \frac{\nu''}{2} + \frac{\nu'^2}{4} + \frac{\mu'+\nu'}{2r}\right],$$

which must hold true for either static or dynamic conditions, and after substituting solution form (8) we get

$$-v''\left[\frac{\mu_0}{2}e^{v/2}+1\right]-\frac{v'^2}{2}\left[\frac{\mu_0}{2}e^{v/2}+1-\frac{\mu_0}{2}e^{v/2}\left(\frac{\mu_0}{2}e^{v/2}+2\right)\right]+$$

$$+v'\left[\left(\mu_1'+\frac{1}{r}\right)\left(\frac{\mu_0}{2}e^{v/2}+1\right)\right]=-\mu_1''+\frac{\mu_1'^2}{2}+\frac{\mu_1'}{r}. \qquad (10)$$

We choose the constant form for $\mu_1$ which leads to

$$v''+\frac{v'^2}{2}\left[1-\frac{\mu_0}{2}e^{v/2}\frac{\left(\frac{\mu_0}{2}e^{v/2}+2\right)}{\left(\frac{\mu_0}{2}e^{v/2}+1\right)}\right]-\frac{v'}{r}=0. \qquad (11)$$

Under quasi-static conditions at time t = 0, we make the substitution $x(r, t=0) \equiv x(r, 0) = \mu_0 e^{v/2}/2$ and get

$$x' = A_0 r(1+x)e^{1+x} \text{ and}$$

$$\ln(1+x)+\sum_{k=1}^{\infty}\frac{[-(1+x)]^k}{k*k!} = A_0\frac{r^2}{2}+A_1. \qquad (12)$$

The metric associated $x$ function, the left side of equation (12), approaches a constant (−0.797) as $x$ approaches zero and interestingly approaches the limit value −C, where C is the "Euler-Mascheroni constant" (0.5772), as $x$ approaches ∞. C is defined as

$$C \equiv \lim_{n\to\infty}\left(\sum_{k=1}^{n}\frac{1}{k}-\ln(n)\right).$$

Examining the large *r* behavior, we demand that the metric $g_{44} = e^\nu$ approach the flat space limit $g_{44} = 1$. For large *r*, *x* approaches minus 1 and the *x* function approaches $ln(1+ x)$, therefore

$$1 + x \approx e^{A_0 \frac{r^2}{2} + A_1}, \text{ or with } A_0 \text{ negative,}$$
$$\mu_0 e^{\nu/2} \approx -2 \text{ and } \mu_0 = -2; \quad x = -e^{\nu/2}. \tag{13}$$

Similarly, $g_{11} = -e^\mu$ is also set equal to $-1$ at large *r*. Since $\mu$ is $\mu_1 + \mu_0 * e^{\nu/2} = \mu_1 - 2 * e^{\nu/2}$, then $\mu_1 = 2$ and $g_{11} = -e^{2(1+x)}$.

At this juncture only the requirement of no radial energy flow and large *r* asymptotic agreement with a flat space form has been demanded. The large r behavior requirement is obviously applicable to the exterior solution, however the character of the space and radiation interior to the sphere in question is not dissimilar to the exterior region. The interior and exterior metrics are therefore considered formally equivalent with only normalization forthcoming.

We now introduce the singular physical requirement, also posed by Tolman in his static description of a constant density sphere embedded in empty space, to uniquely describe the pressure fluctuation, namely that the pressure at the sphere's radius go to zero. We have then a spherical bubble of radiation embedded in a surrounding sea of radiation with the only characterizing distinction being a radial zero in the pressure distribution at the sphere radius.

Reference is now made to equation set (5) where the pressure is given by

$$8\pi T_1^1 = -e^{-\mu}\left[\frac{\mu'^2}{4} + \frac{\mu'v'}{2} + \frac{\mu'+v'}{r}\right] + e^{-v}\left[\ddot{\mu} + \frac{3}{4}\dot{\mu}^2 - \frac{\dot{\mu}\dot{v}}{2}\right] =$$
$$= -8\pi \ pressure\,. \tag{14}$$

We restrict ourselves to the static portion of the equation since in any event we will use the static description as the starting point for the time development. In the $x$ notation we have

$$8\pi \ pressure = e^{-2(1+x)}\frac{x'}{x}\left[x'(x+2) + \frac{2}{r}(x+1)\right]. \tag{15}$$

For the pressure $= 0$ condition, we get

$$0 = e^{-2(1+x)}\frac{A_0 r(1+x)e^{1+x}}{x}\left[A_0 r(1+x)e^{1+x}(x+2) + \frac{2}{r}(x+1)\right]. \tag{16}$$

This equation sets the constant $A_0$ to

$$A_0(x_1, r_1) = -\frac{2}{r_1^2}\frac{e^{-(x_1+1)}}{x_1 + 2} \tag{17}$$

where $x_1 = x(r_1, 0)$ and $r_1$ is the sphere radius. The solution afforded by setting $x_1 = x(r_1, 0) = -1$ is rejected since it leads to requiring $A_1 = \infty$. As mentioned in the discussion of the asymptotic behavior of the metric in equation (13), $A_0$ satisfies the required negative character for the integration constant.

Equation (5) for the energy density $T_4^4$, although at present utilized for a flat space environment, exhibits a curvature energy component. The radiation energy density and this important curvature energy component determine the time evolution of the radiation space, both interior and exterior to the radiation sphere. We rewrite equation (5) in the latter-day expansion factor notation where $e^\mu$ is set equal to $a^2$, $a$ being the expansion factor, to illustrate a comparison to the cosmological form:

$$8\pi T_4^4 = -\frac{1}{a^2}\left[2\frac{a''}{a} - \left(\frac{a'}{a}\right)^2 + \frac{4}{r}\frac{a'}{a}\right] + \frac{3}{1-\ln(a^2)}\left(\frac{\dot{a}}{a}\right)^2 \quad or$$

$$8\pi T_4^4 \equiv "curvature" + \frac{3}{1-\ln(a^2)}\left(\frac{\dot{a}}{a}\right)^2. \qquad (18)$$

This expansion rate equation is to be compared with Friedmann's cosmological expansion rate equation (3). From the definition of $\mu$ and $x$, that is, $\mu = 2(1 + x)$, we also see that since $e^\mu$ is associated with $a^2$, then $d\mu/dt = 2*dx/dt = 2*a^{-1}*da/dt$ and $dx/dt = a^{-1}*da/dt$, the Hubble factor.

In the $x$ notation, the curvature is given by

$$curvature = -e^{-2(1+x)}x'\left[x'\left(\frac{5+3x}{1+x}\right) + \frac{6}{r}\right]. \qquad (19)$$

From the energy density equation of the equation set (5), we can write with the $x$ notation,

$$\left(\frac{\dot{x}}{x}\right)^2 = \frac{8\pi\rho(t)}{3} - \left[-\frac{e^{-2(1+x)}}{3}\left(x'^2\frac{5+3x}{1+x} + 6\frac{x'}{r}\right)\right] =$$
$$= \frac{8\pi\rho(t_{core-birth})}{3}e^{-4(1+x)} - \left[-\frac{e^{-2(1+x)}}{3}\left(x'^2\frac{5+3x}{1+x} + 6\frac{x'}{r}\right)\right]. \quad (20)$$

Equation (20) represents an energy balance statement in that the difference between the radiation and/or matter energy and the warping, or spatial curvature energy, goes into the energy of spatial expansion. Radiation energy density $\rho(t)$ has been represented more explicitly in terms of the time evolution parameter given by $e^{-4(1+x)}$ and subsequently, as an initial condition, will be expressed in terms of the present day radiation energy density, $\Omega r0$. Initial conditions are such that in those regions where the curvature energy density is positive and exceeds the radiation energy density, then $(dx/dt)^2$ is negative and a collapsing space condition will ensue. With an initially uniform radiation energy density and, as a result of the radial behavior of the curvature component, the region internal to the collapse boundary is cut off from the expanding external region since the temporal expansion rate, $dx/dt$, goes to zero at this boundary (zero propagation velocity). In addition, there exists throughout the immediate proximal region of the sphere a variable propagation velocity since the null geodesic produces a velocity equal to

$$propagation\ velocity = \sqrt{\frac{e^\nu}{e^\mu}}. \quad (21)$$

In the static solution of Schwarzschild and the isotropic static solution of Tolman, where $e^\nu = ((1 - r_s/r)/(1 + r_s/r))^2$ and $e^\mu = (1 + r_s/r)^4$, the propagation velocity goes to zero at the Schwarzschild radius $r_s$. As a boundary condition in the present model, the radius of the collapse zone, referred to as the hole region, and the radial zero of the propagation velocity ( at $x(r_{hole}) = -e^{\nu/2} = 0$ ), are set equal thus providing the normalization referred to earlier. This boundary condition determines $A_1$ as

$$A_1 = \ln(1) + \sum_{k=1}^{\infty} \frac{-1^k}{k * k!} - A_0 \frac{r_{hole}^2}{2}. \tag{22}$$

The interior and exterior metrics are equal at the sphere's radial boundary but not equal to the static Schwarzschild solution values, where the exterior solution is for empty space; the empty-space static-solution values are again,

$$g_{44}(r_1) = \left(\frac{1 - r_s/r_1}{1 + r_s/r_1}\right)^2 \text{ and } g_{11}(r_1) = -(1 + r_s/r_1)^4. \tag{23}$$

With the values of $A_0$ and $A_1$ thus determined, $x(r,0)$, moreover, is now provided and, in particular, the value of $x(r_1,0)$ is fixed (see equation (12)) when evaluated at the boundary $r_1$,

$$\ln(1 + x_1) + \sum_{k=1}^{\infty} \frac{[-(1 + x_1)]^k}{k * k!} = A_0(x_1, r_1) \frac{r_1^2}{2} + A_1(x_1, r_1). \tag{24}$$

Figure 1 displays the $x$ function versus the radial coordinate $r$ with the transition from positive to negative at the hole radius. The metric quantity, $x_1$, will be necessary for a subsequent calculation of birth times. The solution for $x_1 = x(r_1,0)$ is

$$\ln(1+x_1) + \sum_{k=1}^{\infty} \frac{[-(1+x_1)]^k}{k*k!} = -\frac{e^{-(x_1+1)}}{r_1^2(x_1+2)}\left(r_1^2 - r_{hole}^2\right) + \ln(1) + \sum_{k=1}^{\infty} \frac{-1^k}{k*k!}. \quad (25)$$

We choose to make the hole mass and thus the hole (collapsed zone) radius an experimentally adjusted parameter of the model. We will, however, subsequently theoretically posit that the formation process is reversible, or quasi-static, and therefore require that, during sphere formation, the entropy change $\Delta S = 0$. This requirement leads to a theoretical determination of the hole radius.

The metrics thereby determined are displayed in Figure 2 as a function of the radial coordinate $r$. We have also included, for reference purposes, the Schwarzschild metric boundary values. Figure 3 illustrates the significant features of the propagation velocity where the asymptotic approach to $c$, at large $r$, is evident along with the zero at the singularity surface. Curvature, temperature and pressure plots are shown in Figures 4, 5 and 6. We have used a hole mass to sphere mass ratio of $5.2(10)^{-3}$ (mass ratio $\equiv (r_{hole}/r_{sphere})^3$; mass energy density equivalent to radiation energy density) in these calculations in anticipation of utilizing an experimentally determined quantity as an input to the model. Curvature exhibits a positive region, at small values of the radial coordinate, transitioning to a negative region and subsequently approaching zero (flat space) at large *r*. The pressure plot displays the zero introduced as a model condition at the sphere boundary and a singularity at the hole boundary caused by the zero metric normalization. Pressure inside the hole region is large and negative and at the hole edge approaches the singularity. Between the hole edge and the sphere radius, the pressure is positive and rapidly decreasing to zero at the sphere radius. External to the sphere, the negative pressure is small and decreases to zero at large radial distances. For some conditions of density and curvature, the hole region increases and when the hole radius approaches the sphere radius, the pressure singularity at the hole edge and the zero of the pressure function at the sphere radius merge at the sphere radius as illustrated in Fig. 7. Figure 8 shows, on a linear scale, the pressure factor near the zero at the sphere's radius.

Again it is instructive to compare these results with the earlier static solutions. In the Schwarzschild treatment, as constructed by Tolman, of a constant density sphere embedded in flat empty space, the curvature energy is zero in the exterior region. Tolman uses the standard form of the spherically symmetric metric and calculates the interior solution to be

$$ds^2 = -e^\lambda dr^2 - r^2 d\Omega^2 + e^\nu dt^2 =$$

$$ds^2 = -\frac{1}{1-\frac{r^2}{R^2}} dr^2 - r^2 d\Omega^2 + \left[ A - B\sqrt{1-\frac{r^2}{R^2}} \right]^2 dt^2$$

$$\text{with } A = \frac{3}{2}\sqrt{1-\frac{r_1^2}{R^2}}, \quad B = \frac{1}{2} \quad \text{and} \quad R^2 = \frac{r_1^3}{r_s}. \tag{26}$$

Although not calculated there, the interior curvature component is

$$curvature = -e^{-\lambda}\left[\frac{\lambda'}{r} - \frac{1}{r^2}\right] - \frac{1}{r^2} = \frac{3}{R^2}. \tag{27}$$

If one assumes that present day galaxy holes derive from such a collapse process and that the present day hole mass (neglecting accretion processes) represents the current collapsed radiation mass, then for the two region model, calculation of the bounds on the galaxy-core birth time are forthcoming from consideration of the galaxy-core mass radius and the hole-mass singularity-radius. If the initial mass distribution's maximum radius at time $t_{core-birth}$ is less than or equal to the singularity radius for that mass, then all of the region will eventually collapse but the collapse process and the mass distribution itself will be subsequently unobservable to regions outside the singularity radius. This case corresponds to the density being equal to the curvature value at the sphere's edge. Model cases with smaller masses or earlier galaxy birth times will additionally contain the second, more rapidly expanding, negatively curved region. Collapse, as used here, describes the reduction of the time metric $x^2 = e^v$ toward zero. The spatial metric, $e^\mu = e^{2(1+x)}$, approaches $e^2$ as $x$ approaches zero. Therefore, for the region internal to the singularity surface, time tends toward zero. The hole then is a region of unchanging spatial character and with no time character.

The time-bound (birth) calculation sets the collapsed mass radius, which equals $r_1 * (hole\ mass\ ratio)^{1/3}$, equal to the sphere radius. This upper time bound is calculated to be $1.5(10)^{-2}$ years, for a galaxy-core mass of $10^9$ stellar masses using equations (30) and (31) for the space expansion factor for a "radiation plus matter" universe and 0.52% for the hole-mass/core-mass ratio (present day mass density $\Omega_{s0}$ relates the galaxy-core radius to the galaxy-core mass). For mass and birth-time combinations greater than the upper extremum, the present day collapsed mass value would be greater than the experimental hole value referred to above. The calculational result for the lower time bound for the galaxy-core birth time (resulting for a galaxy-core mass radius that approaches a singularity radius equal to zero) is $2.7(10)^{-4}$ years. For cosmological times less than this bound, the radiation energy density exceeds the curvature energy density and the spheres experience no collapsing region. This dependence of birth times on galaxy-core mass is illustrated in Fig. 9.

The basic physical concept of causality requires that a coherency time, or formation time, be imposed on the time origin of these structures. Although the radiation energy density can be calculated as a function of time and associated with the time of birth of the radiation spheres, the coherency time interval must additionally be imposed to determine actual birth times. Such a coherency time interval is expressed as

$$t_{coherency} = n_{coherency}\frac{r_1}{c} \ . \tag{28}$$

The coherency time interval is a measure of the time necessary to propagate formation information across the radial extent of the spherical radiation energy distribution and thereby establish equilibrium throughout. Given an instant in time when such structures would have been born, the coherency time interval determines the additional time increment to establish the starting time, or birth time, for subsequently calculating the evolutionary character of the radiation sphere. The uncertainty in expressing such a time is contained in the undetermined constant, $n_{coherency}$. At present no a priori deductive determination of this factor, however, has been forthcoming and so we proceed with the defining equations of the model and merely determine this constant as a companion calculation to be compared for reasonableness with the galaxy-core birth-time calculations. An interpretation of the fact that the experimental hole mass birth-time is so near to the lower time bound birth-time (as shown in Fig. 9) is that the radiation spheres being thus formed require only a minimum coherency time interval to begin the evolutionary contraction (hole region) and expansion (outer shell region) process. The lower time-bound birth-time can be written as

$$t_{core\ birth} = \left(\frac{3}{f0}\right)^{1.5} \frac{GM}{c^3}, \quad where\ f0 = -\left[x'(0)^2 \frac{5+3x(0)}{1+x(0)} + 6\frac{x'(0)}{r(0)}\right]\frac{1}{g_{11}(0)} \approx 2.1,$$

$$or\ t_{core\ birth} \approx 0.85\frac{r_s}{c}\ ;\ x'(0) = -2\sqrt{\frac{g_{44}(r(0))}{g_{44}(r_1)}}\frac{r(0)}{r_1^2}\frac{1+x(0)}{2+x(r_1)}. \tag{29}$$

*f0* is a measure of the curvature function evaluated at the metric zero or hole radius.

The radius of the structure in question is a function of the mass and the density at the time *t* of formation, $t_{core\text{-}birth}$, and is given by equations (30);

$$R_{sphere} = r_1 = \left(\frac{3M_{sphere}}{4\pi\rho(t)\rho_c}\right)^{1/3}, \quad \text{where}$$

$$\rho(t) = \Omega r0 \left(\frac{a_0}{a}\right)^4 = \Omega s0 \left(\frac{3}{4}\frac{age}{t}\right)^2. \quad (30)$$

We have made a distinction between the time of creation and the time of core-birth, where the time of core-birth is equal to the time interval of coherency or formation after the time of creation. The density is changing during this time interval and we use the density at the core-birth time for the model calculations. Expressing Friedmann's equation (5) for the time development of the cosmological expansion factor *a*, for a two species (dust and radiation) flat universe, we write

$$\left(\frac{\dot{a}}{a}\right)^2 = \frac{8\pi\rho_{cosmo}}{3} - \frac{k}{a^2} = \frac{8\pi\rho_c}{3}G\left[\Omega s0\left(\frac{a_0}{a}\right)^3 + \Omega r0\left(\frac{a_0}{a}\right)^4\right] \quad \text{where}$$

$$\rho_c = \frac{3}{8\pi G}H_0^2 \quad \text{and} \quad age = \frac{2}{3}\frac{1}{H_0} \quad \text{or} \quad \rho_c = \left(6\pi G\, age^2\right)^{-1};$$

$$Temperature = T_{\exp}\left(\frac{a_0}{a}\right). \quad (31)$$

The present day values for the radiation energy and the matter energy (approx. total energy) are written as $\Omega r0$ and $\Omega s0$ respectively. In integral form we have

$$\int_{a}^{a_0} \frac{ada}{\sqrt{A+Ba}} = \int_{t}^{age} Cdt \,, where\ A = \Omega r0(a_0)^4,\ B = \Omega s0(a_0)^3\ and\ C = \sqrt{\frac{8\pi\rho_c G}{3}}\,. \quad (32)$$

The solution is

$$t = age \left[ \frac{-\left(2\frac{\Omega r0}{\Omega s0} - \frac{a}{a_0}\right)\sqrt{\frac{\Omega r0}{\Omega s0} + \frac{a}{a_0}} + 2\left(\frac{\Omega r0}{\Omega s0}\right)^{1.5}}{-\left(2\frac{\Omega r0}{\Omega s0} - 1\right)\sqrt{\frac{\Omega r0}{\Omega s0} + 1} + 2\left(\frac{\Omega r0}{\Omega s0}\right)^{1.5}} \right]. \quad (33)$$

$M_{sphere}$ is the mass equivalent of the total radiation energy in the sphere and $\rho_{cosmo}$ is the total cosmological energy density. $\rho_{cosmo} = \rho(t_{core-birth})$ provides the initial density condition for the radiation sphere. We have used the classical notation of equations (2), (3) and (18) for the cosmological space expansion factor *a*. Before time $t_{core-birth}$ the universe is assumed governed by the expansion, space density and temperature behavior determined by this two species characterization, evolving temporally according to the two-component expansion-factor equation (31). At small *a*, or early times, this is approximately,

$$t = age\left(\frac{a}{a_0}\right)^2 \frac{3}{4}\left(\frac{\Omega s0}{\Omega r0}\right)^{1/2}, \quad (34)$$

thereby producing equation (30). At times near the birth times of the structures considered, we have plotted in Fig. 10 the expansion factor *a*, the cosmological density $\rho$ and the radiation temperature. These cosmological values determine the initial conditions for the radiation-sphere time-development equation (20).

The coherency time is now set equal to the time $t$, or $t_{core\text{-}birth}$, of equation (30). Since at the core edge, the time metric factor, $x$, goes to zero, we can use equations (12), (20), (25), (30) and (31) to solve explicitly for $n_{coherency}$ in terms of the remaining undetermined quantity, $r_{hole}$, or $m_{hole}$ ;

$$m_{hole}/M_{sphere} = 5.2(10)^{-3} \; ; \quad x_1 = x_1(r_{hole}/r_1) = -0.579 \; ;$$

$$n_c \equiv n_{coherency} = \frac{1}{4}(x_1+2)e^{x_1+1}\left[-\frac{5}{3}\left(\frac{r_{hole}}{r_1}\right)^2 + (x_1+2)e^{x_1}\right]^{-.5} = 0.626 \, . \, (35)$$

The hole mass to sphere mass ratio is an experimentally measured quantity; Kormendy [5], Magorrian [6], and Kormendy [7] (also see Fig. 2 in Gebhardt [8]). From Magorrian, log ($M_{hole}/M_{bulge}$) = -2.28 (mean with std.dev. = 0.51) or $M_{hole}/M_{bulge} = 5.2 \times 10^{-3}$. We equate our sphere mass to $M_{bulge}$. The coherency constant varies from approximately 0.6 to 2.2 for collapse zone radii from zero to $r_1$. Objectively, these values seem reasonable, thereby supporting the core-birth time values as physically and calculationally defensible.

In Fig. 11 we display the sphere density and curvature and the resultant velocity-factor behavior, as a function of $r$, while incorporating the experimental hole mass value in the calculations. Also shown in Fig. 12 is the character of the velocity-factor solution, at the birth time of the sphere, relative to the background radiation Hubble factor, $Ho^2$. The velocity-factor behavior, mentioned earlier as a measure of the energy balance, shows a resultant overall decrease for the spatial expansion process in the physical vicinity of the sphere. No energy from the embedding space has been provided to the spheres during sphere and hole formation if the entropy condition is satisfied but spatial expansion has slowed.

Rewriting equation (5) in the $x$ notation, after incorporating the equation of state relationship between radiation pressure and radiation density, we have

$$8\pi T_1^1 = -8\pi\, pressure = e^{-\mu}\left[\frac{\mu'^2}{4} + \frac{\mu'\nu'}{2} + \frac{\mu'+\nu'}{r}\right] - e^{-\nu}\left[\ddot{\mu} + \frac{3}{4}\dot{\mu}^2 - \frac{\dot{\mu}\dot{\nu}}{2}\right] =$$

$$= -\frac{8\pi T_4^4}{3} = \frac{e^{-\mu}}{3}\left[\mu'' + \frac{\mu'^2}{4} + \frac{2\mu'}{r}\right] - e^{-\nu}\left[\frac{1}{4}\dot{\mu}^2\right]. \qquad (36)$$

Or, with the aid of the definitions in equations (15) and (19), we get

$$\left(\frac{\ddot{x}}{x}\right) = \frac{8\pi\rho}{3}(1-2x) - pressure\frac{x}{2} - curvature\left(\frac{1}{3} - \frac{x}{2}\right). \qquad (37)$$

Similarly to Figures 11 and 12, we calculate and display in Fig. 13 the results for the acceleration factor, $d^2x/dt^2 + (dx/dt)^2$. Time rate of change of the cosmological background radiation Hubble factor, $dHo/dt$, is compared with the acceleration factor in this figure.

Collapse begins immediately at radii less than the singularity radius. The process is a logarithmic one, however, representing the time integral of equation (20); e.g. a $(10)^6$ fold reduction in $x$ at the central radius requires approximately $7(10)^5$ seconds when calculated with a radiation mass of $(10)^9$ stellar masses and a hole mass ratio of $5.2(10)^{-3}$. The time evolution equation, in integral form, for the metric factor $x$, is given by equation (38);

$$\int_0^t dt = \int_{x0}^{end} \frac{dx}{xc} \left[ \frac{8\pi\rho(t_{core-birth})}{3} e^{-4(1+x)} + \frac{e^{-2(1+x)}}{3}\left( x'^2 \frac{5+3x}{1+x} + 6\frac{x'}{r} \right) \right]^{-1/2}. \quad (38)$$

The $x$ integrand and an approximating function $y(x) = a/x + b$ are plotted in Fig. 14. The approximation is excellent. Integration limits are indicated by $x0$ and $end = (10)^{-6}x0$. In this particular example, we use the radial coordinate at $r = (10)^{-5}r_1$, the experimental hole mass and a sphere mass of $(10)^9$ stellar masses. The energy density function, that is, the sum of the radiation energy density and the curvature energy density, is slowly varying as a function of $x$ and therefore the $x^2$ term dominates the dependence on $x$, thus generating the approximate logarithmic behavior.

Reiterating, the cosmological density, $\Omega s0$ and the hole mass are experimental parameters. For all of the calculations for Figures 4, 6-8 and 11-14, a galaxy-core mass = $10^9$ stellar masses, $\Omega s0 = 1$ and a Hubble's constant $H_0 = 0.485\times10^{-10}$/year (producing an age of $13.7\times10^9$ years) were used. However, if, from an additional theoretical perspective, the radiation sphere formation process

D.R.Koehler    23

behaves preferentially to minimize the entropy change, then an appropriate

governing entropy restriction for zero entropy change can be stated as

$$\Delta S = 0 = \int \frac{dQ}{T} = 2\int \frac{\left(4\pi r^2 e^{3\mu/2} e^{\nu/2} dr\right)}{T_0 e^{-\mu/2}} \left(-e^{-2(1+x)} x' \left[ x'\left(\frac{5+3x}{1+x}\right) + \frac{6}{r} \right]\right). \tag{39}$$

We have used Tolman's expression (eqn. 97.2) for the total energy of the radiation sphere utilizing the perfect fluid or radiation character to express $T_1^{\,1}$, $T_2^{\,2}$ and $T_3^{\,3}$ in terms of $T_4^{\,4}$ and expressing $T_4^{\,4}$ as the curvature energy from equation (5). The energy of formation, that is the curvature energy, is integrated over the entire spatial domain and the temperature is a function of the spatial metric $e^{-\mu/2}$; $T = T_0 e^{-\mu/2}$. A numerical integration of equation (39) yields a theoretically determined value for $x_1$ and therefore $r_{hole}$. The result is $r_{hole} = 0.72\ r_1$. At this point, it should be noted again that the gravitating radiation sphere is a rather diffuse structure where there is no radiation density difference across the pressure = 0 boundary at $r_1$ nor does the curvature or distortion function exhibit any discontinuity across this same boundary. The time evolution of the sphere, deriving from spatial expansion, preserves the main spatial qualities of the sphere but it is apparent that the comparison with the experimental hole-mass to bulge-mass ratio is somewhat arbitrarily defined. The ratio is calculated to vary from $(0.72)^3$ with $r_{sphere} \equiv r_1$ to $(0.72/6)^3 = 1.7(10)^{-3}$ with $r_{sphere} \equiv 6r_1$ (see Fig .4 for the radial extent of the curvature function). Matter accretion processes occurring after sphere formation would more probably be responsible for bulge mass formation. The time-evolved radiation sphere itself then should be interpreted as the "hole structure".

What is the thermodynamic distribution function describing the expected number of radiation spheres of a given mass, or given energy content? A partitioning of the total available energy with domination by the self, or rest-mass, energy of the spheres was considered appropriate. A gas based analogy and an associated Maxwell Boltzmann distribution where

$$\delta n = nAM^3 e^{-E/kT} dE \quad \text{with} \quad E = Mc^2 \quad \text{and}$$

$$A = c^{-2} \left( \int_0^\infty e^{-Mc^2/kT} M^3 dM \right)^{-1} = c^{-2} \left( \frac{c^2}{kT} \right)^4 (\Gamma(4))^{-1} \quad \text{leading to}$$

$$\delta n = \frac{n}{6} \left( \frac{E}{kT} \right)^3 e^{-E/kT} \frac{dE}{kT} \qquad (40)$$

does not, however, seem an adequate description. The energy peak in this distribution occurs at $E_{max} = 3kT$. At this time, though, the desired distribution function has not been formulated.

In chronological review, the presence at time $t_{creation}$ of an incompletely formed pressure fluctuation begins the departure from relative uniformity, which describes the early universe. A time $t_{core\ birth}$ is required for coherency or formation of the radiation spheres and is followed by both a collapsing and expanding time evolution of the distorted spatial region. In the present radiation, or perfect fluid, modeling, the description provided by equations (5) incorporates only the gravitational physics of the collapsing galaxy-core space; any subsequent matter accretion processes, or other energy sources, are not included. The continuous matter- but radiation-dominant-energy of the early universe, that constitutes the galaxy-core material, is considered the perfect fluid used to determine the expansion factor solution form. A present day experimental hole mass value, equal to 0.52% of the galaxy bulge-mass (we have associated the galaxy bulge-mass with our core-mass), has been used for the evolving hole although we have introduced an alternative theoretical entropy concept which constrains the determination of the hole radius. This radius is assumed to be the metric singularity value. The hole radius during and after the logarithmic collapse does not change while the time character of the hole collapses to zero. Temporal evolution of the galaxy-core expansion parameter, in the outer regions of the energy distribution, follows the negative curvature behavior and, asymptotically at large r, a flat (curvature = 0) behavior. Since the spatial extent of the hole region does not change, the galaxy-core density function in the hole region remains constant during the evolution. The variable $T_{exp} = T_{experimental}$ is the cosmic microwave background temperature and is utilized as the reference. Temperature is assumed to follow an $a^{-1}$ or $e^{-\mu/2}$ behavior throughout both spatial domains.

If we calculate the radiation energy and birth time relationship for these radiation spheres (see eqn.(29)) near the cosmological inflation time, we get approximately $(10)^3$ equivalent grams (for the hole mass) at $t_{core\ birth} = (10)^{-33}$ seconds ( $(10)^{-33}$ seconds is defined as the end of inflation ). In this calculation, we have used the hole mass rather than the sphere mass. The contemporary attempt to relate a collapsed gravitational entity to the fundamental particle spectrum, for example baryons, therefore is seen in the present model to lead to the requirement of an additional mass reduction factor (going backward in time) of $(10)^{-27}$. Since, for these radiation spheres, the mass (from eqn. 29) is inversely proportional to the square root of the density and in the radiation environment we are considering, the density is inversely proportional to the fourth power of the expansion factor (inversely proportional to the second power of the time), a change in the expansion factor proportional to the square root of the mass factor achieves the desired reduction. A change of $(10^{-27})^{.5} = 3.2(10)^{-14}$ ( fourteen orders of magnitude) in the expansion factor therefore would produce the conditions for baryon-like radiation spheres. This puts such a proton-like particle creation-time during the period of cosmological inflation since the inflation period produces an expansion factor change of the order of thirty orders of magnitude. However if the release of latent energy from the phase transition of the inflation process maintained the energy density of the universe during the transition, then further expansion factor reduction would have to have occurred before inflation. For the time period from $10^{-35}$ to $10^{-43}$ seconds (Planck time), however, sufficient reduction could not have resulted. By contrast, in the rest-energy based Maxwell Boltzmann distribution of equation (39), distributions of radiation spheres with holes approximating baryon masses would peak ($3kT$) at a temperature of approximately $7.1(10)^{14}$K which occurs at about $6.6(10)^{-10}$ seconds. At the post inflation time of $(10)^{-33}$ seconds, the temperature is $5.8(10)^{26}$K representing a peak sphere mass of $2.7(10)^{-13}$ grams and an associated hole mass of $1.4(10)^{-15}$ grams.

## 3. OBSERVATIONAL REDSHIFT MODELING

For radiation emission and detection processes, photon propagation behavior along the time evolving emitter to observer path, is determined by the equation for null geodesics and integration over time along the light path. This leads to the expression relating the emitted and observed time intervals, Linder [11],

$$\int_{t_e}^{t_{e2}} dt/a = \int_{t_o}^{t_{o2}} dt/a \; ; \; t_{e2} = t_e + \Delta t_e \; ; \; t_{02} = t_0 + \Delta t_0 \; . \tag{41}$$

The resulting Friedmann-Robertson-Walker (FRW) redshift expression as in Linder [11], is

$$z = \frac{\nu_e - \nu_o}{\nu_o} = \Delta t_o / \Delta t_e - 1 = a(t_o)/a(t_e) - 1 \tag{42}$$

and illustrates that only the initial and final expansion states, $a_j(t_i)$ (i = emitter time or observer time, j = emitter or observer), determine the overall resultant change in frequency or wavelength of the propagating radiation. In other words, in a non-monolithic universe where local warping is present and where radiation emission sources and radiation observers (detectors) are both embedded in such locally warped regions, calculation of the radiation modification (redshift) involves calculation of the local region's expansion state as manifest in the local expansion parameter $a_j(t_i)$. The potential energy, or wavelength, diagram illustrated in Fig. 15 is a pictorial representation of the evolving photon energy state as it propagates (1), through the emitting warped galaxy region, (2), out of the warped interface, (3), through the spatially expanding path between emitting and observing galaxies, (4), into the observer space-galaxy interface, and finally, (5), to the detection point within the observer galaxy. The observer galaxy and the emitter galaxy are assumed to exhibit the same time evolutionary or expansion characteristics. In such a path, the wavelength stretching (photon energy loss) step at the emitter galaxy-space interface and the energy loss process during intergalactic travel is mirrored at the second observer space-galaxy interface where the energy loss is partially recovered and the photon wavelength decreases. The expanding emitter and observer galaxy regions therefore produce the net overall energy change, or photon wavelength increase, during the time interval from emission to detection. If the $a_{emitter}$ and $a_{observer}$ evolution lines corresponded to the same $t^{2/3}$ time behavior as $a_{space}$, then no energy loss or recovery would be incurred at the space-galaxy interfaces. Although the present model calculations are limited to the core expansion-factor time-development, the notion of local space warping and its impact on the propagating radiation is still appropriate.

If the galaxy environment influences radiation redshifting, then microwave background radiation (CBR) is also affected and should display a wavelength offset equivalent to the ratio of expansion parameter values between present-day observer galaxy-space and intergalaxy-space;

$$CBR\ offset \equiv \left(\frac{a_{observer}(r, today)}{a_{space}(today)}\right) = \left(\frac{e^{\mu/2}(r, today)}{a(today)}\right) \ ;$$
$$Actual\ temp_{CBR} = Measured\ temp_{CBR} * CBR\ offset \ . \tag{43}$$

A radiation evolution and propagation diagram is shown in Fig. 16 and, as illustrated, suggests an actual inter-galaxy background radiation temperature lower (longer wavelength) than measured inside the galaxy. However, since $a_{space}(today)$ is probably greater than that in the outer regions of the galaxy, a less-than-one CBR offset is predicted for observers in these regions. That is, appropriate galaxy models, with observers in high mass-density regions, would predict a smaller intergalaxy CBR temperature than that which is measured, a phenomenon that derives from the local space warping produced by matter in the vicinity of the detector. The outer regions of the cores, or hole-bulge masses, described here display a rapidly decreasing radial warping (see Fig. 4), however, and, when considered as emission sources, would apparently cause measurable impact on galactic spectrometric electromagnetic frequency shifts only when the emissions emanated from regions very close to the cores.

## 4.   SUMMARY

In conclusion, we have modeled radiation generated galaxy cores, born in early time frames, exhibiting collapse zones with black hole type characteristics. The galaxy-cores, as modeled, display radius-dependent curvature, pressure and expansion rates and exhibit time evolution rates approaching the background $t^{1/2}$ dependencies in the outer regions while the inner regions are collapsing. We also postulate from this modeling that cosmological redshift data are interpretable as measurements of a localized galactic expansion parameter, both at the emitter and at the observer, and that cosmic microwave background radiation measurements should be impacted by the difference between galaxy and intergalaxy expansion rates.

## 6. FIGURE CAPTIONS

FIG. 1. Metric x-factor diagram as a function of the spherical radial coordinate (in units of the radiation sphere radius, $r_1$) illustrating the zero at the hole radius $h$.

FIG. 2. Model metrics: the spatial, $g_{11,2} = -g_{11} = e^{\mu}$, and the temporal, $g_{44,2} = g_{44} = e^{\nu}$, metrics are illustrated as a function of the radial coordinate $r$ (in units of the radiation sphere radius, $r_1$); the subscript 2 indicates usage of the experimental hole-mass value. The Schwarzschild radius, S, and the static-empty-space solutions, G11(negative) and G44, are shown for reference.

FIG. 3. Propagation velocity as a function of the radial coordinate $r$ (in units of the radiation sphere radius, $r_1$).

FIG. 4. Curvature, on a linear scale, as a function of the radial coordinate $r$ (in units of the radiation sphere radius, $r_1$). Curvature is in units of $(10)^{-30}/cm^2$.

FIG. 5. Two dimensional representation of the three dimensional curvature function. Negative values of the displayed function represent positive curvatures.

FIG. 6. Pressure and temperature (K) throughout the radiation sphere's proximity. The radial coordinate is in units of the radiation sphere radius $r_1$. Pressure is in units of $(10)^{-40}/cm^2$. The reference value, $(10)^{10}$, is indicated.

FIG. 7. Pressure function versus radius for the case of the hole mass equal to the experimental value (subscript 1) and for the hole radius equal to the sphere radius (subscript 3) where the singularity at the hole boundary has merged with the zero at the sphere radius. Radial coordinate in units of the sphere radius $r_1$.

FIG. 8. Pressure in the immediate proximity of the radius of the sphere. A linear pressure scale showing the behavior at the pressure zero. The radial coordinate is in units of the radiation sphere radius $r_1$.

FIG. 9. Radiation sphere birth times (years) versus mass of the radiation sphere. T_Bir0 is for a zero hole-radius extremum, T_Bir2 is from the experimental hole-mass datum and T_Bir1 for a hole-radius equal to the sphere radius. Radiation sphere mass values extend from $10^9$ to $10^{12}$ stellar masses.

FIG. 10. Cosmological radiation and matter (dust) densities and temperature as a function of time (years) in the early universe. The dominance of the radiation energy density over the matter energy density at these early times is noted. Cosmological temperatures, multiplied by a factor of $(10)^{20}$ for display purposes, are near $10^8$ K while the radiation energy density is $1.4(10)^{27}\ \rho_c$ at $2.7(10)^{-4}$ years.

FIG. 11. Metric rate change (*X_DOT2*), curvature (*KURV*) and energy density factor (*ΩRAD*), all on a logarithmic scale, are displayed as a function of the radial coordinate in units of the sphere radius $r_1$. The hole radius, *h*, and the Schwarzschild radius, *S*, are indicated. *ΩRAD* is $2.9(10)^{-30}/\text{cm}^2$. The crossover

from positive curvature dominance to radiation energy dominance creates the hole region near the radial origin.

FIG. 12. Metric rate change, $(dx/dt)^2 = X\_dot2$, on a linear scale, showing comparison to the background Hubble factor $H_0^2 (\text{sec}^{-2})$.

FIG. 13. Metric-rate rate-change, $d^2x/dt^2 + (dx/dt)^2 = X\_ACC$ on a logarithmic scale, showing comparison to the background Hubble-factor rate-change $dH_0/dt = Ho\_dot$. The Schwarzschild radius, $S$, is indicated

FIG. 14. Metric rate change integral. The $x$ function integrand, from equation (38), is shown over an $x$ range from "$x0$" = $x1s = x_{start}$ to "end" = $(10)^{-6} x_{start}$. In this example the integrand is accurately approximated by the $y(x)$ function $y(x) = a/x + b$, with $a = 2.1(10)^4$ and $b = -4(10)^4$.

FIG. 15. Potential Energy or Wavelength Diagram for photon propagation along a galaxy-emitter to galaxy-observer path.

FIG. 16. Potential Energy or Wavelength Diagram for photon propagation along a microwave-emitter to galaxy-observer path.

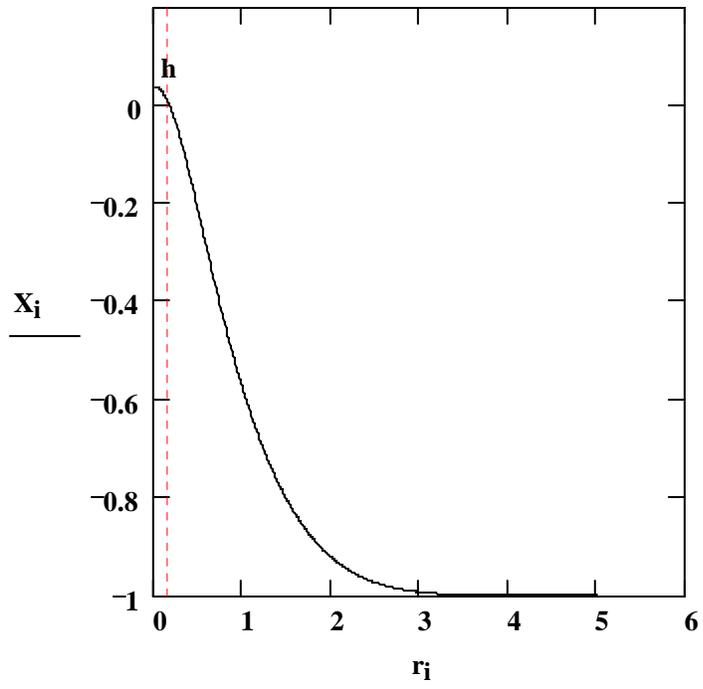

Figure 1    Dale R. Koehler

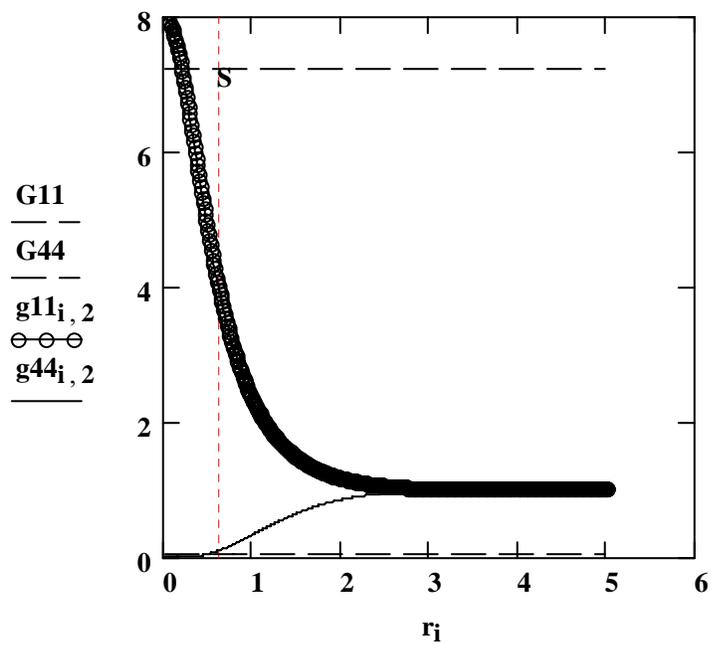

Figure 2  Dale R. Koehler

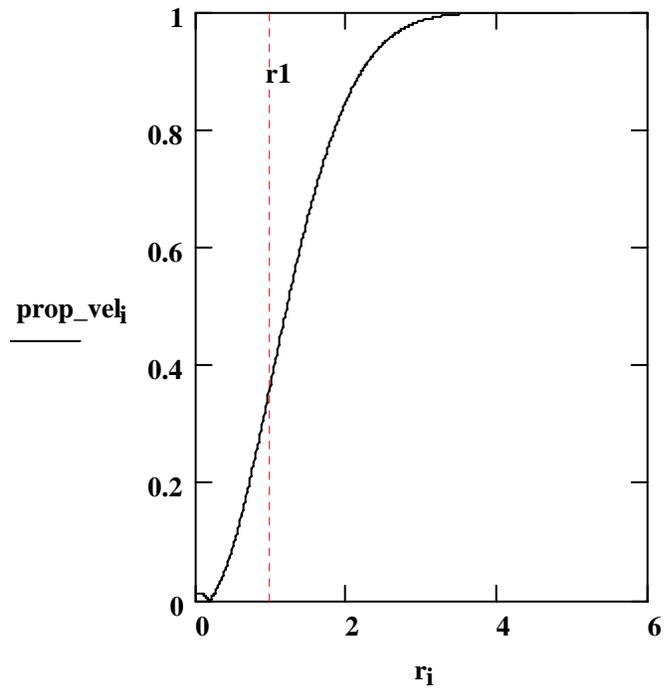

Figure 3     Dale R. Koehler

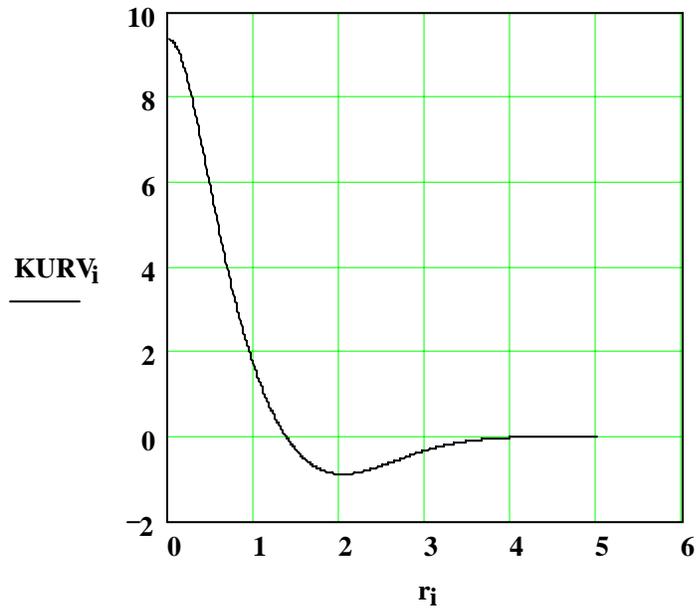

Figure 4    Dale R. Koehler

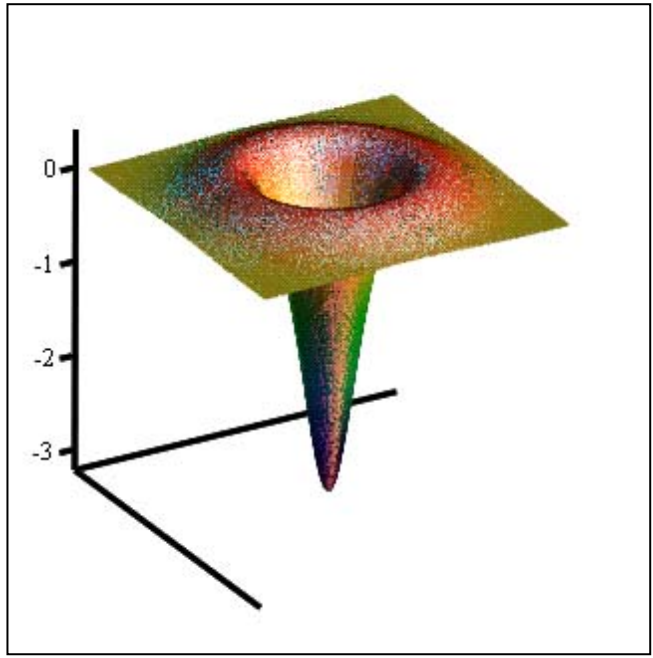

**CURV**

Figure 5   Dale R. Koehler

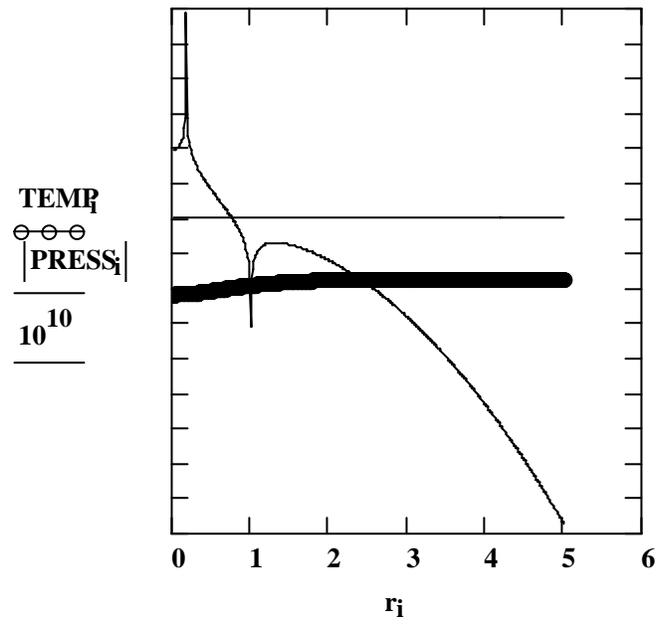

Figure 6    Dale R. Koehler

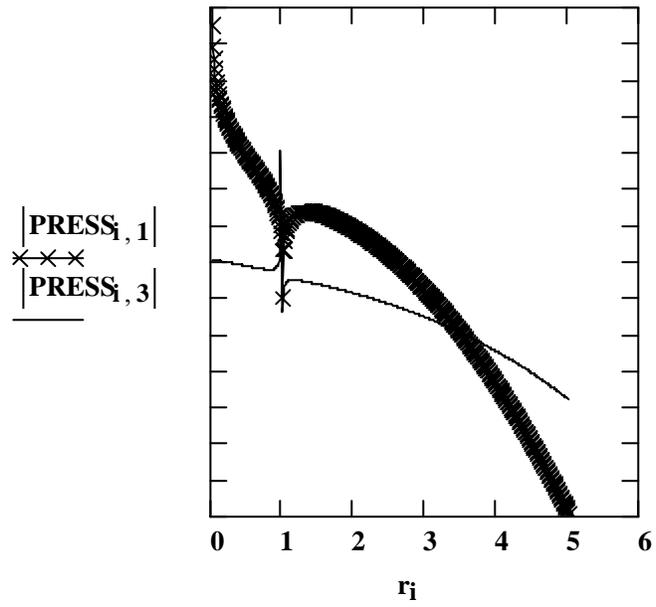

Figure 7      Dale R. Koehler

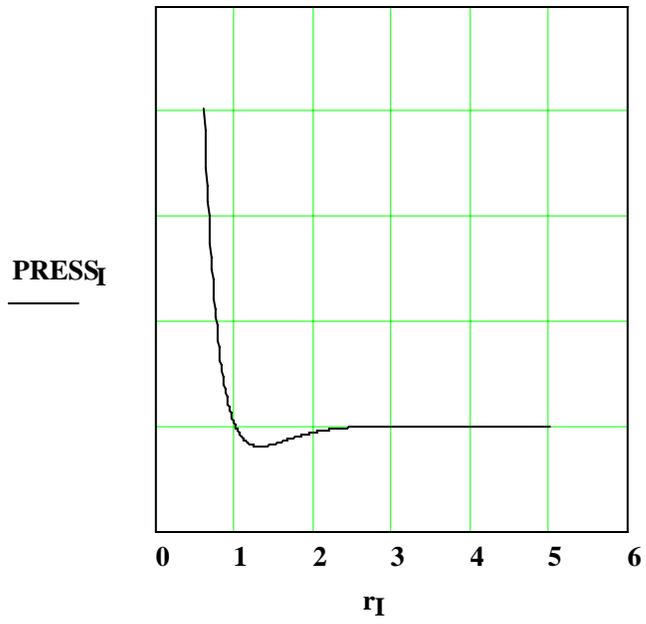

Figure 8    Dale R. Koehler

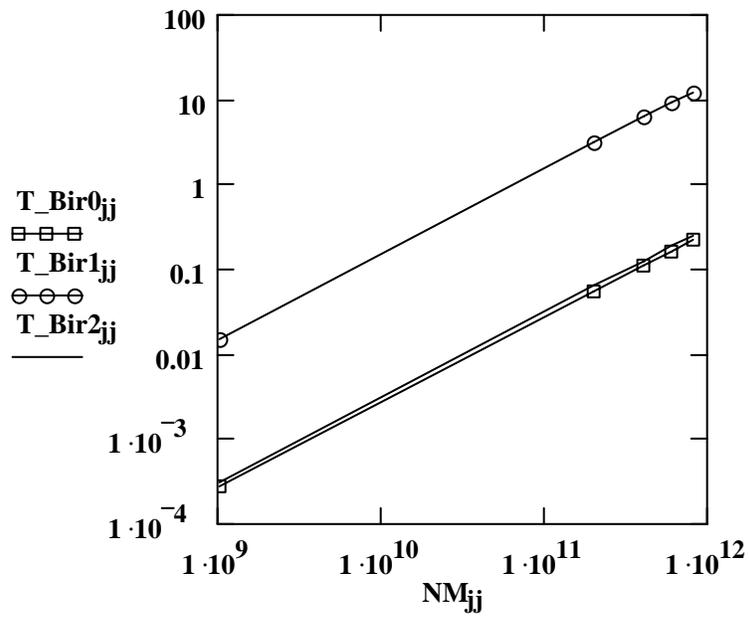

Figure 9  Dale R. Koehler

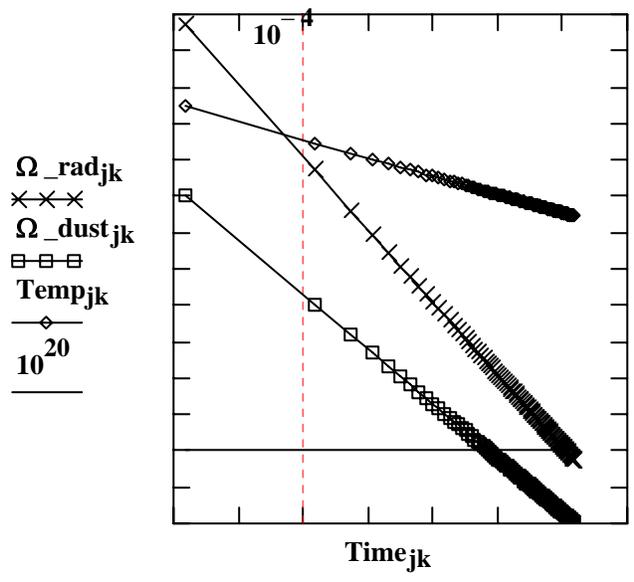

Figure 10   Dale R. Koehler

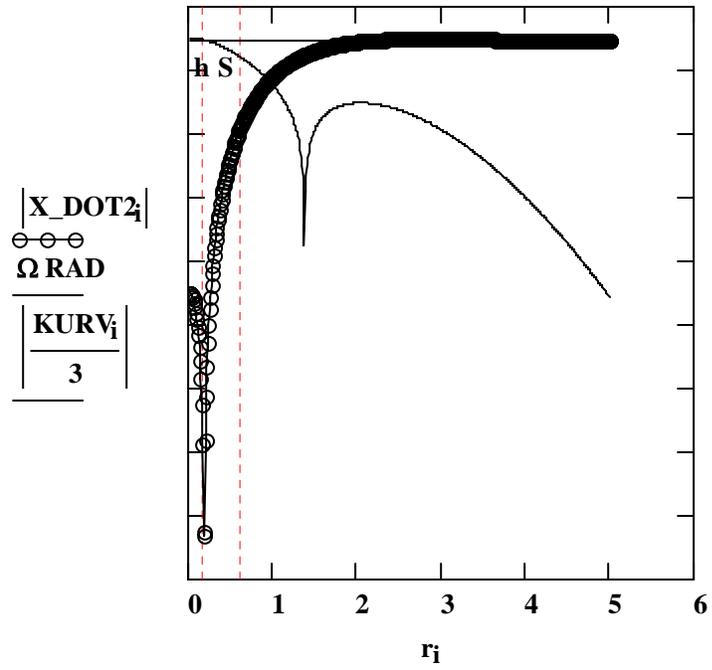

Figure 11     Dale R. Koehler

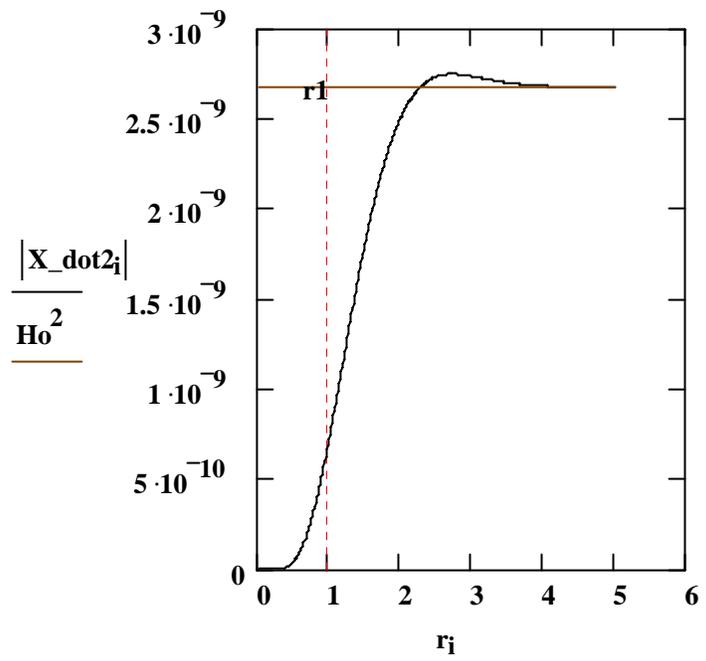

Figure 12    Dale R. Koehler

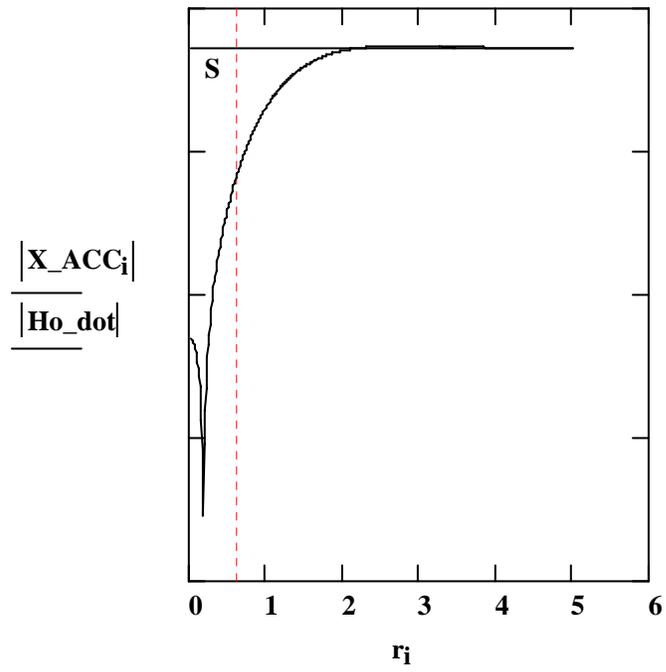

Figure 13     Dale R. Koehler

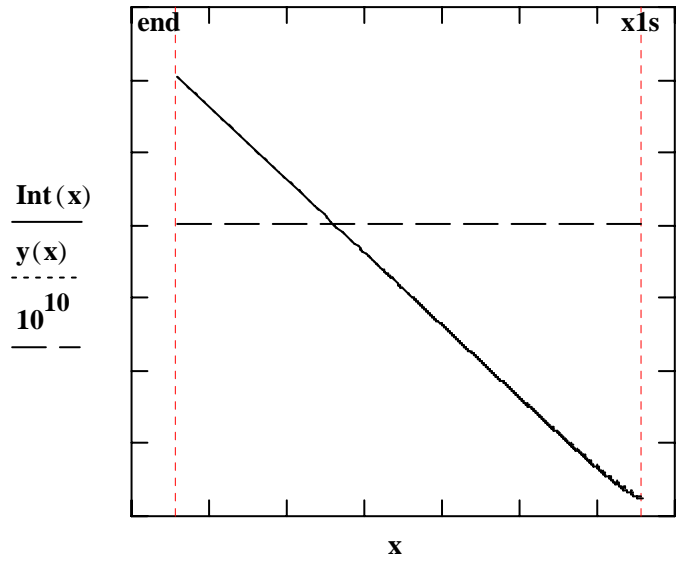

Figure 14  Dale R. Koehler

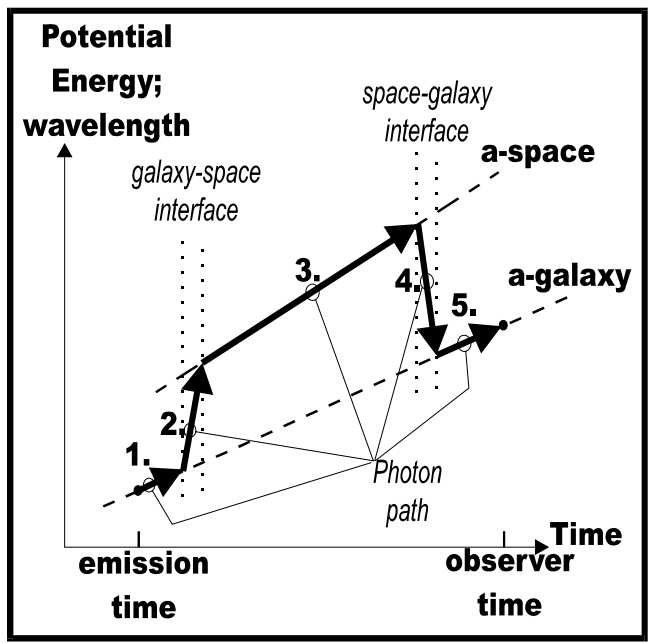

Figure 15     Dale R. Koehler

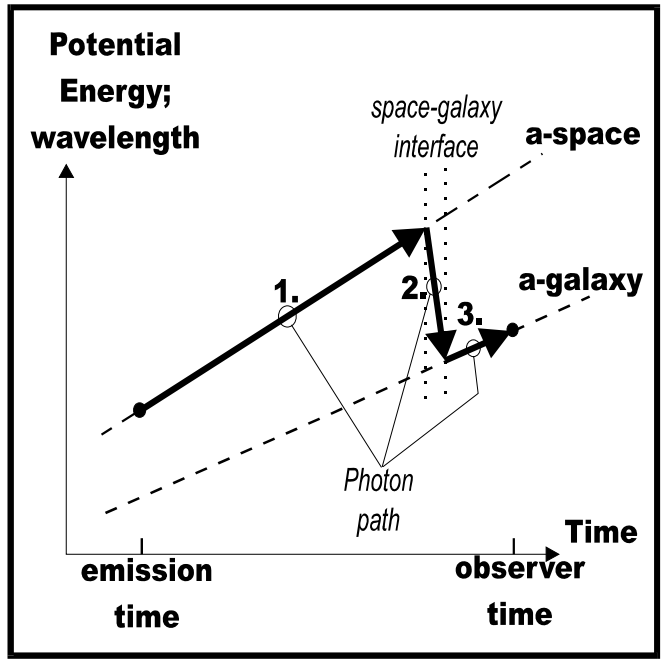

Figure 16    Dale R. Koehler